\documentclass{article}

\usepackage{arxiv}

\usepackage[utf8]{inputenc} 
\usepackage[T1]{fontenc}    
\usepackage{hyperref}       
\usepackage{url}            
\usepackage{booktabs}       
\usepackage{amsfonts}       
\usepackage{nicefrac}       
\usepackage{microtype}      
\usepackage{lipsum}
\usepackage{graphicx}
\usepackage{multirow}
\graphicspath{ {./images/} }
\usepackage{siunitx}
\usepackage{amsmath}

\newcommand{\beginsupplement}{%
    \setcounter{table}{0}
    \renewcommand{\thetable}{S\arabic{table}}%
    \setcounter{figure}{0}
    \renewcommand{\thefigure}{S\arabic{figure}}%
 }

\title{Breast cancer histopathology image-based gene expression prediction using spatial transcriptomics data and deep learning}

\author{
 Md Mamunur Rahaman \\
  School of Computer Science and Engineering\\
  University of New South Wales\\
  Kensington, NSW 2052, Australia \\
  \texttt{md\char`\_mamunur.rahaman@unsw.edu.au} \\
   \And
 Ewan K. A. Millar \\
  Department of Anatomical Pathology\\
  NSW Health Pathology, St. George Hospital\\
  NSW 2217, Australia \\
  \And
 Erik Meijering \\
  School of Computer Science and Engineering\\
  University of New South Wales\\
  Kensington, NSW 2052, Australia \\
}

\begin{document}
\maketitle
\begin{abstract}
Tumour heterogeneity in breast cancer poses challenges in predicting outcome and response to therapy. Spatial transcriptomics technologies may address these challenges, as they provide a wealth of information about gene expression at the cell level, but they are expensive, hindering their use in large-scale clinical oncology studies. Predicting gene expression from hematoxylin and eosin stained histology images provides a more affordable alternative for such studies. Here we present BrST-Net, a deep learning framework for predicting gene expression from histopathology images using spatial transcriptomics data. Using this framework, we trained and evaluated 10 state-of-the-art deep learning models without utilizing pretrained weights for the prediction of 250 genes. To enhance the generalisation performance of the main network, we introduce an auxiliary network into the framework. Our methodology outperforms previous studies, with 237 genes identified with positive correlation, including 24 genes with a median correlation coefficient greater than 0.50. This is a notable improvement over previous studies, which could predict only 102 genes with positive correlation, with the highest correlation values ranging from 0.29 to 0.34.
\end{abstract}


\section{Introduction}
\label{s:introduction}
Breast cancer is the most prevalent malignancy in women and is a molecularly diverse disease. In 2018, 2.1 million women were diagnosed with breast cancer, one every 18 seconds, and 626,679 women died, a 3.1\% yearly rise~\cite{Harbeck2019}. Early breast cancer, where the disease is limited to  the breast with or without the involvement of the axillary lymph nodes, has a good prognosis with a 5-year survival of close to 90\%. However, advanced metastatic disease is often incurable with existing therapies, which aim to delay progression and treat symptoms. According to the World Health Organization (WHO), early identification of cancer significantly improves the likelihood of appropriate decision making and successful treatment~\cite{Krithiga2021}.

Pathological analysis of a tissue biopsy removed from the breast tumour is acknowledged as the gold standard method for the diagnosis, subtype classification and staging. Heterogeneity is evident at the morphological and molecular levels of breast cancer, where divergent patterns of gene expression and interaction with host immune and stromal populations can vary widely within and between similar tumour types. A better understanding of spatial gene expression is a current area of very active cancer research to better capture spatial differences in gene expression which may more accurately reflect patient outcome and response to treatment, which are currently difficult to gauge.

Emerging spatial transcriptomics (ST) technologies enable profiling gene expression at a single-cell resolution while preserving spatial orientation and cellular composition within a tissue~\cite{Stahl2016, Larsson2021}. ST is quickly becoming an extension of single-cell RNA sequencing (scRNAseq). To understand the complex transcriptional architecture of biological systems, it is necessary to determine how cells are arranged in space and how gene expression changes in different foci of a targeted tumour or tissue. ST methods based on next generation sequencing (NGS), such as $10\times$ Genomics' Visium, Slide-Seq~\cite{Rodriques2019}, Slide-Seq2~\cite{Stickels2020}, and HDST~\cite{Vickovic2019}, barcode whole transcriptomes but have restricted capture rates and resolutions greater than a single cell (around $100~\mathrm{\text\textmu m}$ for Visium and $10~\mathrm{\text\textmu m}$ for Slide-Seq). Moreover, NGS-based approaches offer unbiased profiling of large tissue sections without requiring a list of target genes, unlike image-based technologies such as in-situ hybridization (ISH) and in-situ sequencing (ISS)~\cite{Ke2013, Young2020, Rao2021}. In order to accurately and robustly analyse the data produced by ST technologies, specialised computational methods are required because of the data's intrinsic noise, high dimensionality, sparseness, and multimodality (containing histology pictures, count matrices, etc.).

Although ST provides a plethora of data, its generation is prohibitively costly, precluding its large-scale application. In addition, with commercial equipment like the $10\times$ Genomics Visium system, significant expertise is required to create high-quality expression patterns for whole tissue samples~\cite{Monjo2022}. On the other hand, compared to ST, hematoxylin \& eosin (H\&E)-stained histology images are not only easier and less expensive to acquire, but they are also routinely used in clinical practice. Therefore, the recent development of predicting spatial gene expression data from routine histology images offers advantages in cost, speed, and availability of such data. These projections have the potential to provide virtual ST data, which will make it possible to investigate the regional differences in gene expression on a large scale.

While existing analysis pipelines mostly use gene expression profiles, not image pixel values, the integration of imaging data with gene expression data is a new area of research~\cite{Tan2020}. Moreover, whole-slide images (WSIs) have been utilised to predict global gene expression patterns via HE2RNA~\cite{Schmauch2020}, demonstrating their strong correlation with transcription. HE2RNA is trained to predict gene expression patterns from The Cancer Genome Atlas (TCGA) WSIs without specialist interpretation. Similarly, ST-Net uses ST data in conjunction with DenseNet to make predictions on the spatially varying gene expression of each spot over WSIs~\cite{He2020}. Prediction of the spatially resolved transcriptome makes it possible to use images to look for biomarkers that change in various tumour location spots.

Although these techniques have performed well, they do have certain drawbacks. HE2RNA lacks the capacity to learn from ST data since it was designed for bulk RNA sequencing. Despite the fact that ST-Net is purpose-built for ST, its convolutional neural network (CNN) model is pretrained on the ImageNet dataset, which fails to take into account the observed individual spatial locations of spots. Expression of genes generally has local patterns, thus it is vital to consider spatial location when training networks to make accurate predictions.

To overcome these challenges, we have developed BrST-Net (short for Breast ST-Net), a framework based on deep learning that uses spatial transcriptomics data to predict gene expression directly from breast histopathology images (Fig.~\ref{workflow}). This framework includes stain normalisation of tissue sections, filtering of gene and ST spots, and the generation of image patches depending on the locations of the spots. The image patches are then utilised to train CNN and transformer models, with the primary network devoted to predicting 250 highly expressed genes. We introduce an auxiliary network before the end of the main network to forecast the remaining genes, which ultimately enhances the generalisation performance of the main network by functioning as an alternative kind of regularisation~\cite{Palacio2020}. We have thoroughly trained and evaluated 10 recent cutting-edge deep learning models to assess their prediction capabilities on the ST dataset. Our suggested approach outperforms existing work: the framework using EfficientNet-b0 with an auxiliary network can predict 237 genes with positive correlation, whereas ST-Net can only predict 102 genes with positive correlation. ST-Net identified the top-5 predicted genes in their framework with a median correlation coefficient value of 0.34, 0.33, 0.31, 0.30, and 0.29, while our framework using EfficientNet-b0 with an auxiliary network can predict 24 genes with a median correlation coefficient greater than 0.50, which is a substantial performance boost.

\section{Materials and methods}
\label{s:materials}

\subsection{Dataset description}
We used the fifth edition of the publicly accessible spatial transcriptomics dataset \cite{Stenbeck2021} as also used by ST-Net \cite{He2020}. It comprises a total of 68 WSIs of H\&E-stained frozen tissue sections, from 23 breast cancer patients having various subtypes of breast cancer, including luminal A, luminal B, triple-negative, human epidermal growth factor receptor 2 (HER2) luminal, and HER2 non-luminal. Each section is scanned at 20$\times$ magnification, with image size approximately 9,500 $\times$ 9,500 pixels in JPEG format, and contains around 450 spots, each having a diameter of 100 µm and approximately 3,000 gene expression counts. In total, the dataset comprises 30,612 spots, and includes spot coordinates, count matrices, and coordinate files.

\subsection{Preprocessing of the dataset}

\subsubsection{Stain normalization}
Histology samples stained with H\&E often vary in colour between and within laboratories from one batch to the next. Stain normalisation is employed as a preprocessing step in computational pathology algorithms to regulate variances and has a significant impact on the outcome of automated image analysis methods~\cite{Ciompi2017}. In our work, we employed StainTools \cite{Peter554_StainTools}, using the Vahadane technique to normalize the stain variance in the target image and the Luminosity Standardizer function to standardize the brightness, both of which facilitate consistent interpretation of the images \cite{Vahadane_StructurePreserving_2016}.

\begin{figure*}[!t]%
\centering
\includegraphics[width=\textwidth]{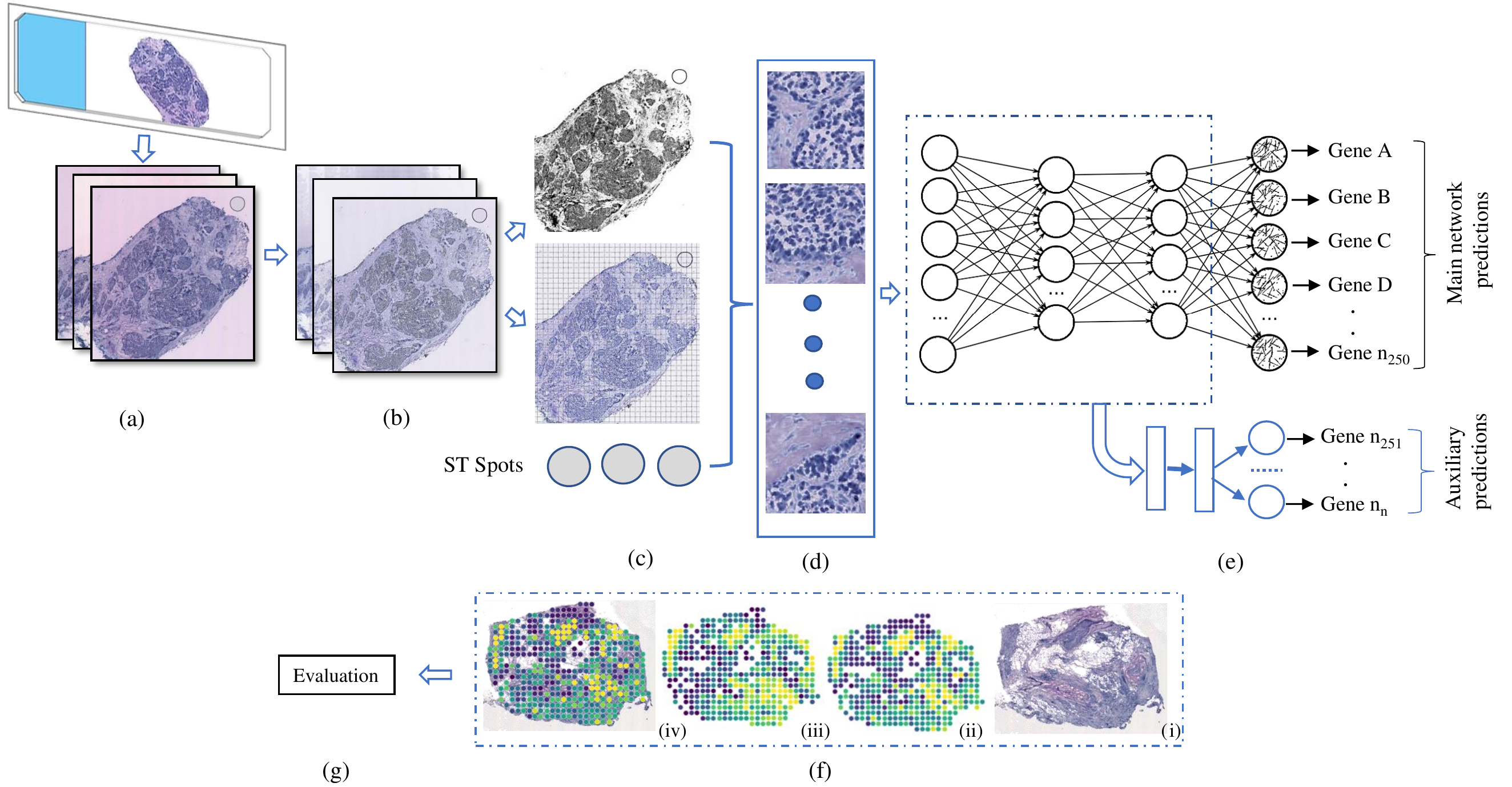}
\caption{Overview of the proposed BrST-Net framework. (a) Original whole slide image. (b) Stain normalised image. (c) Mask image and filtered ST spot location information. (d) Extracted image patches using the spot location information. (e) Main network with an auxiliary network for gene expression prediction. (f) Prediction of an individual gene on the tissue sample, showing (i) test sample, (ii) ACTB gene prediction, (iii) ground-truth expression of gene ACTB, and (iv) gene ACTB overlayed on the tissue sample. (g) Performance evaluation.}
\label{workflow}
\end{figure*}

\subsubsection{Gene and spots filtering}
To eliminate background noise from gene expression data per spot, we removed genes with a mean expression of zero across all samples, and selected spots with at least 1,000 total read counts for analysis. The resulting filtered genes were stored in a ZipFile format (.npz) comprising multiple arrays, including count, pixel, patient, and index arrays. This resulted in approximately 28,792 spot files from all samples, which were then utilized to generate corresponding image patches. After filtering, approximately 6,000 genes remained. The complete preprocessing pipeline is described in detail in~\cite{Seal2013}.

\subsubsection{Generating image patches}
Since the original images are too large for direct input to a CNN, we extracted $N$ patches from the WSIs centered on the ST spots, using their size and position. The patches were then flattened into an $N \times (3 \times m \times n$) matrix, where $m$ and $n$ are the patch's width and height, respectively, and each patch has 3 colour channels. In our experiment, a value of $m=n=224$ pixels was selected. To ensure that the training dataset only contained informative data, we excluded patches with more than 50\% white pixels, as these are unlikely to contain relevant information and may introduce noise in the data. In total, 27,652 image patches were acquired for training.

\subsubsection{Gene symbol conversion}
The dataset provides a comprehensive list of genes, including their Ensembl identifiers (IDs), which are widely used as a standard for gene annotation and identification. The IDs were translated from the given format to gene symbols according to the HUGO Gene Nomenclature Committee (HGNC) database (\href{genenames.org}{genenames.org}) and converted into a Pickle file, retaining just the Ensembl ID and its related symbol. Pickle is a Python library to store complex data as binary files.

\subsubsection{Data augmentation}
Histology image analysis is supposed to be orientationally invariant, just like a pathologist can analyse a microscopic image from any angle. Therefore, to facilitate the generalisability of the trained networks, we used PyTorch's ``torchvision.transforms'' module to randomly transform image patches during training. It performs random horizontal and vertical flipping, and random rotation of 90 degrees to increase the diversity of the training data. All patches were converted to a PyTorch tensor and normalised to zero mean and unit variance.

\subsection{Gene expression prediction models}
We experimented with 10 state-of-the-art CNN and transformer models as the basis for our BrST-Net framework (Fig. \ref{workflow}), including ResNet, InceptionNet, six variants of EfficientNet, and two vision transformer models, briefly described below. These main networks were trained to learn spatial features from histology images and predict gene expression levels. However, due to the high dimensionality of ST data, CNN architectures may be limited in their ability to predict gene expression levels accurately for the remaining genes. To address this limitation, we modified the main network by introducing an auxiliary network (AuxNet) just before the end of the network. The AuxNet is a simple feed-forward neural network with a single fully connected layer that predicts the remaining genes. It enhances the generalisation performance of the main network and helps prevent overfitting, providing an alternative form of regularization. 

\subsubsection{ResNet model}
ResNet (Residual Network)~\cite{He_2016_CVPR} is a deep learning architecture widely used in computer vision tasks such as image classification, object detection, and semantic segmentation \cite{9216455, Jiang2019}. It uses residual connections to improve information propagation and learning, expressed as $F(x) = H(x) + x$, where $x$ is the input, $H(x)$ is the learned residual mapping, and $F(x)$ is the residual block output. This helps mitigate the vanishing gradient problem in deep neural networks. In this study, ResNet101 is used as the main network and fully trained on the ST dataset, making this the first study to use ResNet for gene expression prediction.

\subsubsection{Inception model}
Inception \cite{Szegedy2015} is a deep convolutional neural network designed for image classification. It uses a combination of convolutional and pooling layers, along with auxiliary classifiers, to capture both local and global features in the input image. The final prediction is made by combining the outputs of the main and auxiliary classifiers, represented mathematically as $y = Wx + b$, where $x$ is the input feature vector, $W$ is the weight matrix and $b$ the bias term, and $y$ is the output. Different versions exist, with Inception-v3 achieving state-of-the-art performance on benchmark datasets and now being widely used in real-world applications~\cite{AlHusaini2022, Saini2019}. In this study, we fully trained it on the ST dataset and evaluated its performance for gene expression prediction.

\subsubsection{EfficientNet model}
The conventional approach to increasing the learning capacity of CNN models involves tweaking the network's depth $d$ and width $w$ and the image resolution $r$. While this enhances accuracy, it often requires significant manual tuning and can result in suboptimal performance. This has been addressed in literature by examining various scaling strategies and proposing a systematic network architecture scaling method, leading to the development of EfficientNet \cite{Tan2019}. The scaling method employs a user-defined coefficient, $\varphi$, to scale up networks in a more systematic manner, according to the following equations:
\begin{equation}
\begin{array}{c}
	d = \alpha^\varphi, \quad w = \beta^\varphi, \quad r = \gamma^\varphi, \\[1em]
    \text{s.t.} \quad \alpha\cdot\beta^2\cdot\gamma^2 \approx 2, \quad \alpha \geq 1, \quad \beta \geq 1, \quad \gamma \geq 1,
\end{array}
\end{equation}
where $\alpha$, $\beta$, and $\gamma$ are calculated through automatic hyperparameter optimisation using grid search. The scaling of the model is regulated by $\varphi$, which controls the allocation of computational resources. In pathological image analysis, multiple studies have utilized the EfficientNet model and achieved superior accuracy~\cite{Wang2021, Kallipolitis2021, Ahmad2022, Byeon2022, Munien2021}. To our knowledge, however, our study is the first to use EfficientNet for gene expression prediction using the ST dataset. Specifically, we experimented with versions b0, b1, b2, b3, b4, and b5.

\subsubsection{Vision transformer model}
Recent editions of the ImageNet Large-Scale Visual Recognition Competition have witnessed the dominance of the vision transformer (ViT) over other state-of-the-art approaches. Inspired by the challenge of natural language processing (NLP) to deal with varying sentence lengths, ViT aims to deal with dependencies at various spatial distances, and partitions an image into a fixed number of patches. In the feature extraction stage of a ViT model, a 2D image $x \in \mathbb{R}^{X \times Y \times C}$ of size $X \times Y$ pixels and $C$ channels, is transformed into a 1D sequence $x_{p} \in \mathbb{R}^{N \times P^2 \times C}$ of $N$ patches of size $P^2$, where $N$ is equal to the input sequence length of the transformer encoder and is calculated as $N = XY/P^2$. In the transformer, all patches are flattened and then projected linearly to $D$ dimensions, referred to as patch embeddings. These patch embeddings are the input to the transformer encoder, preserving positional information~\cite{Dosovitskiy2020}. The transformer encoders consist of multiple multihead self-attention (MHSA) and multilayer perceptron (MLP) blocks. The MHSA layer in the transformer is capable of learning the attention for spots or image patches and involves a linear combination of several attention heads:
\begin{equation}
\text{MultiHead}(Q, K, V) = [\text{head}_1,\dots,\text{head}_c] \times W_0,
\end{equation}
where $c$ denotes the number of heads, $W_0$ are the weights used to aggregate the attention heads, and $Q$, $K$, and $V$ represent the query, key, and value, respectively. Each head is calculated as:
\begin{align}
    \text{head}_i &= \text{Attention}\!\left(QW_i^Q, KW_i^K, VW_i^V\right), \\
    \text{Attention}(Q, K, V) &= \text{softmax}\!\left(\frac{QK^T}{\sqrt{d_k}}\right) V,
\end{align}

where $W_{i}^{Q}$, $W_{i}^{K}$, $W_{i}^{V}$ are weight matrices. Here, $QK^T/\sqrt{d_K}$ refers to the attention map, whose pattern is $N \times N$, and $V$ represents the self-attention mechanism's value, where $V = Q = K$. The attention mechanisms produces an $N \times 1024$ matrix as output. In our experiment, we evaluated the performance of the ViT-B16 and ViT-B32 models, where B16 and B32 refer to the partitioning of an image into 16 and 32 patches, respectively \cite{Dosovitskiy2020}.

\subsection{Model training and evaluation}

\subsubsection{Loss function}
To train a network, its weights are iteratively updated to minimise a given loss function. In our framework, we use the following loss:
\begin{equation}
\mathcal{L} = \mathcal{L}_\text{main} + \lambda \cdot \mathcal{L}_\text{aux},
\end{equation}
where $\mathcal{L}$ is the overall loss, $\mathcal{L}_\text{main}$ is the loss of the main network, $\mathcal{L}_\text{aux}$ is the loss of the auxiliary network, with both losses being calculated using cross-entropy, and $\lambda$ is a hyperparameter used to balance the contribution of the two losses. We experimented with various values of $\lambda$ and found that a value of 40 gives good performance. Stochastic gradient descent (SGD) optimization was used to minimise the loss.

\subsubsection{Error measures}
To quantify errors between predicted and actual values, we used the mean absolute error (MAE) and root mean squared error (RMSE):
\begin{equation}
    \text{MAE} = \frac{1}{n} \sum_{i=1}^{n} |y_i - \hat{y}_i|,\quad
    \text{RMSE} = \sqrt{\frac{1}{n} \sum_{i=1}^{n} (y_i - \hat{y}_i)^2},
\end{equation}
where $n$ is the total number of observations, $y_i$ is the true value of observation $i$, and $\hat{y}_i$ is its predicted value. Both error measures yield an average prediction error of a model, ranging from $0$ to $\infty$, with smaller values indicating more accurate predictions.

\subsubsection{Correlation measure}
To assess the reliability of gene expression predictions from histopathology images, we used the Pearson correlation coefficient (Pcc) \cite{pearson1901, Benesty2009, Onik2018}:
\begin{equation}
	\text{Pcc} = \frac{\sum_{i=1}^{n} (a_i-\bar{a})(b_i-\bar{b})}{\sqrt{\sum_{i=1}^{n} (a_i-\bar{a})^2 \sum_{i=1}^{n} (b_i-\bar{b})^2}},
\end{equation}
where $a_i$ and $b_i$ are the true and predicted genes, respectively, $n$ is the total number of genes, $\bar{a}$ is the mean of the $a_i$, and $\bar{b}$ is the mean of the $b_i$. The Pcc evaluates the linear relationship between two variables and assigns a score ranging from $-1$ to $1$. A score of $1$ means there is a perfect positive linear correlation, a score of $-1$ indicates a perfect negative linear relationship, and a score of $0$ implies there is no linear relationship. In practice, a Pcc score between 0.5 to 1 indicates strong correlation, a score between 0.3 to 0.5 indicates medium correlation, and a score between 0.1 to 0.3 indicates weak correlation.

\begin{table}[!t]
\centering
\caption{Top 10 predicted genes by our framework using different models with AuxNet. The numbers are median Pcc values on the held-out test patient ID BC23903. Here, to save space, ``EfficientNet'' is abbreviated to ``ENet'', and ``Inception-v3'' to ``Incep-v3''. Bold indicates best performance (highest Pcc) per gene and underlined indicates second-best.}
\footnotesize
\resizebox{1\textwidth}{!}{
\begin{tabular}{lccccccccccc}
\hline
\textbf{Genes} & \textbf{ResNet101} & \textbf{Incep-v3} & \textbf{ENet-b0} & \textbf{ENet-b1} & \textbf{ENet-b2} & \textbf{ENet-b3} & \textbf{ENet-b4} & \textbf{ENet-b5} & \textbf{ViT-B16} & \textbf{ViT-B32} \\
\hline
B2M & 0.4729 & 0.5720 & \textbf{0.6325} & 0.5659 & 0.5722 & 0.5199 & \underline{0.6221} & 0.1800 & 0.5475 & 0.3909 \\
ACTG1 & 0.5457 & 0.5391 & \textbf{0.6233} & 0.5637 & 0.6054 & 0.5888 & \underline{0.6220} & 0.2259 & 0.6075 & 0.5231 \\
ACTB & 0.6066 & 0.4985 & \underline{0.6204} & 0.3723 & 0.4843 & 0.5919 & 0.6147 & 0.2718 & \textbf{0.6359} & 0.5452 \\
TMSB10 & 0.3896 & 0.4678 & \textbf{0.6197} & 0.0634 & 0.4597 & 0.5621 & \underline{0.5717} & 0.2038 & 0.5576 & 0.1359 \\
GNAS & 0.5019 & 0.5896 & \underline{0.6139} & 0.5864 & 0.5886 & 0.5976 & \textbf{0.6170} & 0.2588 & 0.5872 & 0.4148 \\
PTMA & 0.5521 & 0.5072 & \underline{0.5902} & 0.5641 & 0.5654 & 0.5423 & \textbf{0.5956} & 0.3685 & 0.5639 & 0.4762 \\
PTPRF & 0.4927 & 0.5269 & \textbf{0.5926} & 0.5667 & 0.5524 & 0.5359 & \underline{0.5568} & 0.0733 & 0.5126 & 0.4643 \\
ERBB2 & 0.3156 & 0.5033 & \underline{0.5897} & 0.3600 & 0.3499 & \textbf{0.5930} & 0.5750 & 0.1684 & 0.5343 & 0.4626 \\
PRDX1 & 0.4561 & 0.5329 &\textbf{0.5702} & 0.4478 & 0.5246 & 0.5405 & \underline{0.5466} & 0.1213 & 0.5262 & 0.4789 \\
TMSB4X & 0.4238 & 0.4884 & 0.5665 & 0.3876 & 0.4203 & \textbf{0.5876} & \underline{0.5872} & 0.2086 & 0.5717 & 0.5345 \\
\hline
\end{tabular}
}
\label{tab1}
\end{table}

\begin{figure}[!b]
\centering
\includegraphics[scale=0.8]{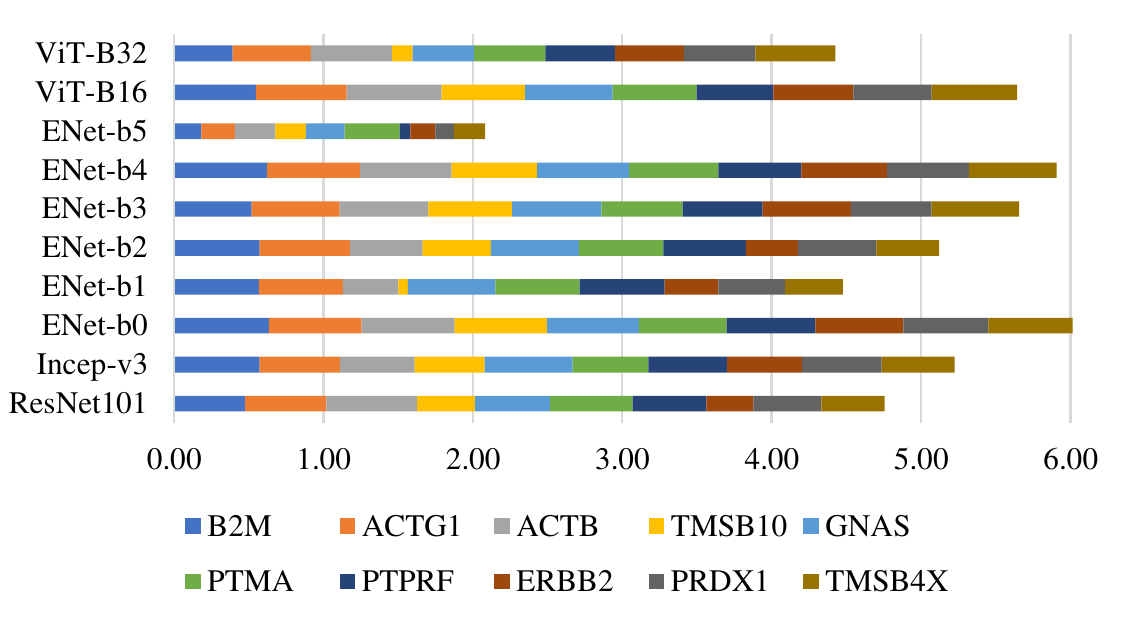}
\caption{Stacked bar chart summarising the predictive performances of the different models.}
\label{stack}
\end{figure}

\section{Experimental results}

\subsection{Implementation and setup}
BrST-Net was implemented in PyTorch and Python 3.7. For training, the batch size was set to 32, the number of epochs to 200, the gene filter to 250, and the learning rate for SGD was set to 0.001. The models were trained on 2 GPUs with 24 CPU cores and 50 GB of RAM on the Gadi cluster of the National Computational Infrastructure in Australia. A 5-fold cross-validation was performed on 22 out of 23 patients to evaluate the performance of the trained models. The remaining patient was held out as a final test set to assess the generalisability of the models.

\subsection{Quantitative results}
\begin{figure}[!t]
\centering
\includegraphics[scale=0.8]{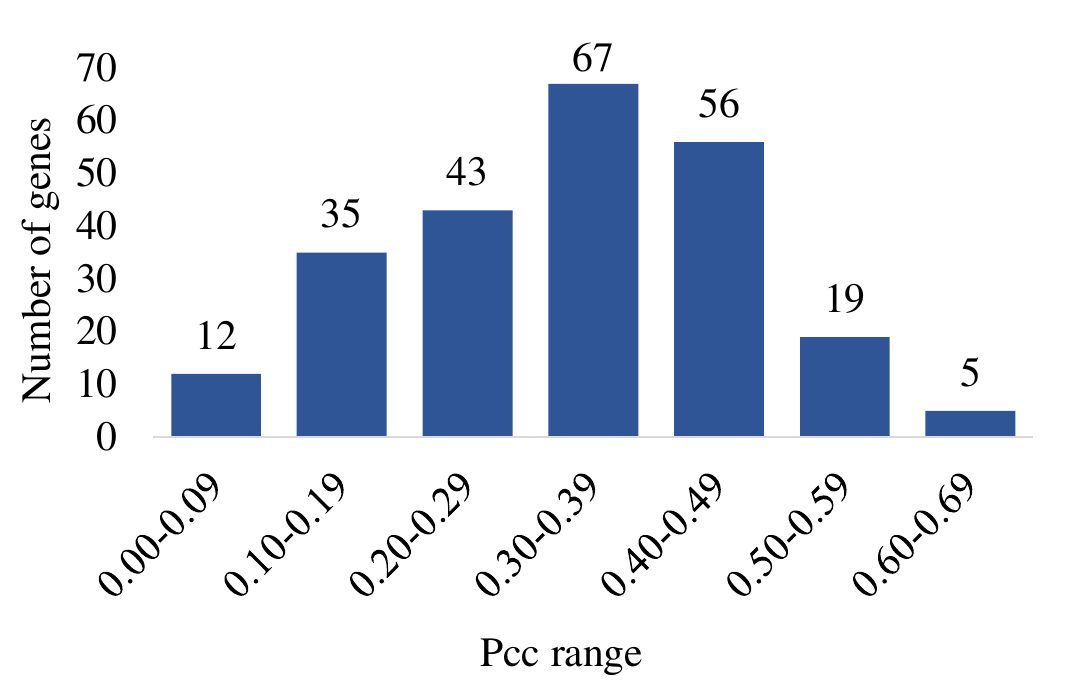}
\caption{Gene prediction Pcc value histogram of the 237 positive correlations produced by EfficientNet-b0 + AuxNet.}
\label{analysis}
\end{figure}

To assess the efficacy of the different models for our framework, we fully trained and evaluated a total of 10 models on the ST dataset: ResNet101, Inception-v3, EfficientNet-b0, EfficientNet-b1, EfficientNet-b2, EfficientNet-b3, EfficientNet-b4, EfficientNet-b5, ViT-B-16 and ViT-B-32, with and without the use of AuxNet, and compared their performances in predicting the top 250 genes. From the results with AuxNet on the held-out test case (Table~\ref{tab1} and Fig.~\ref{stack}), we observe that the EfficientNet architecture is dominant, with EfficientNet-b0 performing most favourably, followed by EfficientNet-b4. The results without using AuxNet for the test case (Supplementary Table S1) clearly show the improvements brought by the proposed auxiliary network. Specifically, for each gene, the best median Pcc value is higher with the use of AuxNet than without, though this higher number may be produced by a different main network model. Of the other models, ResNet101, Inception-v3, and ViT-B16 performed somewhat comparably (no consistent winner), and better than ViT-B32 and EfficientNet-b5, the latter of which generally performed worst. EfficientNet-b0 with AuxNet was able to predict 237 genes with positive correlation (Fig.~\ref{analysis}), including 24 genes with median Pcc values greater than 0.5 (strong correlation), 123 genes with Pcc values between 0.3 and 0.5 (medium correlation), and 78 genes with Pcc values between 0.1 and 0.3 (weak correlation).

\begin{table}[!b]
\centering
\caption{Average mean absolute error (aMAE) and average root mean squared error (aRMSE) of different models with and without AuxNet on the training and testing data.}
\footnotesize
\begin{tabular}{l@{\hspace{2em}}cccc}
\hline
\multirow{2}{*}{\textbf{Model}} & \multicolumn{2}{c}{\textbf{Training Data}} & \multicolumn{2}{c}{\textbf{Testing Data}} \\ \cline{2-5} 
& \textbf{aMAE} & \textbf{aRMSE} & \textbf{aMAE} & \textbf{aRMSE} \\
\hline
ResNet101 + AuxNet & 0.7823 & 0.9654 & 0.7609 & 0.9344 \\ 
Inception-v3 + AuxNet & 0.7831 & 0.9650 & 0.7403 & 0.9036 \\
EfficientNet-b0 + AuxNet & 0.7730 & 0.9528 & 0.7490 & 0.9144 \\
EfficientNet-b1 + AuxNet & 0.7489 & 0.9259 & 0.7345 & 0.8976 \\
EfficientNet-b2 + AuxNet & 0.7618 & 0.9399 & 0.7277 & 0.8898 \\
EfficientNet-b3 + AuxNet & 0.7858 & 0.9677 & 0.7481 & 0.9135 \\
EfficientNet-b4 + AuxNet & 0.8037 & 0.9899 & 0.7718 & 0.9409 \\
EfficientNet-b5 + AuxNet & 0.8042 & 0.9904 & 0.7905 & 0.9716 \\
ViT-B16 + AuxNet & 0.7909 & 0.9754 & 0.7262 & 0.8893 \\
ViT-B32 + AuxNet & 0.8039 & 0.9909 & 0.7699 & 0.9403 \\
\hline
ResNet101 & 0.7830 & 0.9675 & 0.7484 & 0.9158 \\
Inception-v3 & 0.7801 & 0.9657 & 0.7470 & 0.9148 \\
EfficientNet-b0 & 0.7456 & 0.9242 & 0.7499 & 0.9148 \\
EfficientNet-b1 & 0.7529 & 0.9324 & 0.7462 & 0.9122 \\
EfficientNet-b2 & 0.7610 & 0.9419 & 0.7397 & 0.9042 \\
EfficientNet-b3 & 0.8008 & 0.9878 & 0.7465 & 0.9120 \\
EfficientNet-b4 & 0.7635 & 0.9446 & 0.7445 & 0.9100 \\
EfficientNet-b5 & 0.7761 & 0.9586 & 0.7374 & 0.9017 \\
ViT-B16 & 0.7878 & 0.9730 & 0.7621 & 0.9324 \\
ViT-B32 & 0.7961 & 0.9829 & 0.7607 & 0.9311 \\
\hline
\end{tabular}
\label{tab3}
\end{table}

The average MAE and average RMSE of all models with and without AuxNet (Table~\ref{tab3}) show that for the best-performing model according to the Pcc metric (Table~\ref{tab1}) the errors on the testing data are smaller with the use of AuxNet than without. More broadly, Inception-v3, the lower-version EfficientNet models, and ViT-B16 all produce smaller errors with the use of AuxNet. Interestingly enough, the errors of these models on the training data (Table~\ref{tab3}) as well as on the validation data (Supplementary Tables S2 and S3) can be larger with the use of AuxNet. This may be explained by the fact that AuxNet introduces additional outputs in the network, which can make the optimization process more difficult. During training, the network learns to optimize the main objective while also optimizing the auxiliary objectives.  Consequently, this may result in a more complex optimization problem, and the network may not converge as quickly or as well as without the use of AuxNet. However, the use of AuxNet can still enhance the generalisation performance of the models on the unseen testing data, which is crucial in practical applications. Thus, the slight increase in training error is a reasonable trade-off for the improved generalisation performance on the testing data.

\subsection{Visualisation of gene expression}
\begin{figure}[!t]
\centering
\includegraphics[width=\textwidth]{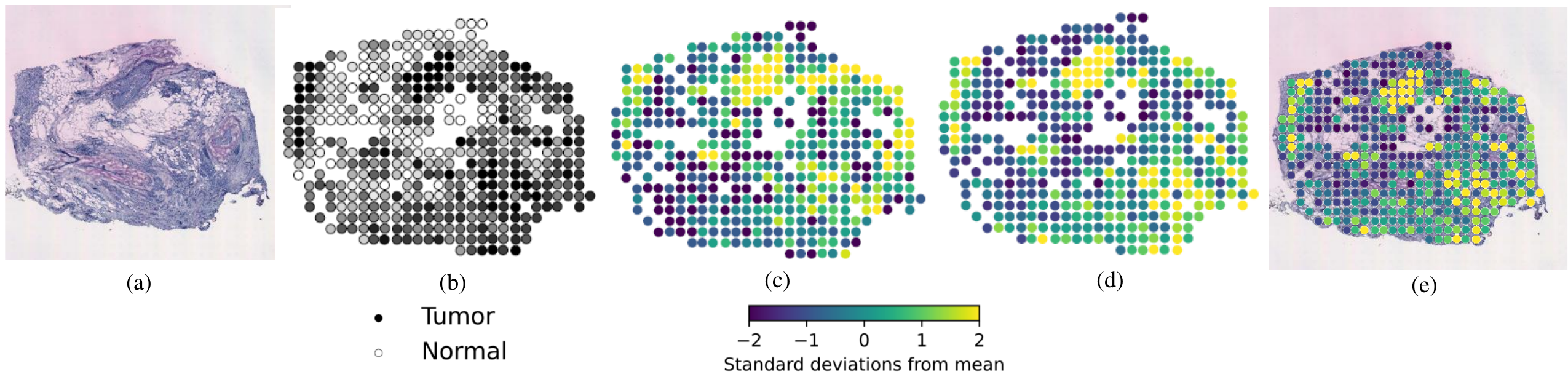}
\caption{Visualisation of gene B2M expression for patient BC23903 slide number C2. (a) Histopathology sample from the test data. (b) Binary label of tumour (black) and normal (white) areas, with gray parts indicating areas that are not purely tumor or normal. (c) Ground truth expression of gene B2M. (d) Predicted expression of gene B2M (Pcc = 0.6325). (e) Visualisation of predicted gene B2M expression overlayed on the tissue slice.}
\label{visualization}
\end{figure}

To better understand the gene expression predictions, the top predicted genes with high Pcc values were visualised on the corresponding histology image (Fig.~\ref{visualization} and Supplementary Figs.~S1 and S2). The visual comparison of the true and predicted expressions of gene B2M and their Pcc values for the test dataset (Fig.~\ref{visualization}) shows a strong correlation between the two. The yellow portions correspond to regions where B2M is strongly expressed, while the blue portions correspond to regions where that gene is poorly expressed. We observe that the yellow portion of the tissue is highly correlated with the presence of the black tumour annotation, indicating that this gene is colocated with cancer cells in the image. This demonstrates the effectiveness of our proposed framework BrST-Net in predicting local gene expression from tissue images for a selected panel of genes.

\subsection{Computational costs}
To evaluate the computational cost of training our framework, we recorded the training times of all considered models in our experiments. From the results (Table~\ref{tab4}) we observe that as EfficientNet increased in scale from b0 to b5, the training time gradually increased accordingly, as expected. ResNet101 and Inception-v3 were about as costly as the lower-scale versions of EfficientNet, while ViT-B16 was about as expensive as the higher-scale versions of EfficientNet, and ViT-B32 was the fastest network. For a few models, the use of AuxNet added comparatively little to the total cost, and for most models, it even slightly reduced the cost, demonstrating that AuxNet is a lightweight network that can improve accuracy while incurring minimal computational overhead. For all models, while training takes many hours, their application at test time takes only a few seconds per image, making them computationally very feasible in clinical practice.

\begin{table}[!t]
\centering
\caption{Computational cost of our framework using different models.}
\footnotesize
\begin{tabular}{l@{\hspace{2em}}c}
\hline
\textbf{Model} & \begin{tabular}[c]{@{}c@{}}\textbf{Training Time}\\ \textbf{(hours:minutes:seconds)}\end{tabular} \\
\hline
ResNet101 + AuxNet & 23:17:50 \\
Inception-v3 + AuxNet & 20:04:46 \\
EfficientNet-b0 + AuxNet & 19:04:28 \\
EfficientNet-b1 + AuxNet & 23:54:54 \\
EfficientNet-b2 + AuxNet & 23:30:49 \\
EfficientNet-b3 + AuxNet & 25:02:51 \\
EfficientNet-b4 + AuxNet & 29:58:24 \\
EfficientNet-b5 + AuxNet & 37:02:11 \\
ViT-B16 + AuxNet & 33:09:25 \\
ViT-B32 + AuxNet & 17:08:06 \\
\hline
ResNet101 & 23:58:59 \\
Inception-v3 & 20:05:58 \\
EfficientNet-b0 & 19:49:15 \\
EfficientNet-b1 & 23:07:10 \\
EfficientNet-b2 & 22:35:53 \\
EfficientNet-b3 & 25:00:06 \\
EfficientNet-b4 & 30:02:30 \\
EfficientNet-b5 & 37:36:46 \\
ViT-B16 & 32:18:28 \\
ViT-B32 & 17:18:28 \\
\hline
\end{tabular}
\label{tab4}
\end{table}

\section{Discussion}
\label{s:discussion}
Spatial transcriptomics technologies can profile gene expression for complete transcriptomics at almost single-cell resolution, with spatial positions matched with H\&E-stained histological images (WSIs). While WSIs are inexpensive, accessible, and commonly generated in clinics, generating ST data is very costly and complicated, and currently, only a limited number of research centres are capable of doing so. Therefore, predicting gene expression directly from WSIs is useful.

In this paper, we proposed BrST-Net, a framework for predicting gene expression using ST data. We fully trained and tested 10 different deep-learning models for our framework to predict 250 genes based on ST breast cancer data. Of all considered models, the combination of EfficientNet-b0 and our proposed AuxNet was able to predict 237 genes with positive median correlation, including 24 genes with strong correlation (Pcc value over 0.5), 123 genes with medium correlation (Pcc values between 0.3 and 0.5), and 78 genes with weak correlation (Pcc values between 0.1 and 0.3). EfficientNet-b4 and EfficientNet-b3 also performed relatively well, as did Inception-v3 and ViT-B16, while EfficientNet-b5 performed worse and ViT-B32 generally did poorly. The fact that, among the transformer models, ViT-B16 outperformed ViT-B32 in gene expression prediction, may be attributed to the smaller dimensionality of ViT-B16, which enhances its ability to capture complex gene expression relationships. These findings emphasize the significance of selecting an appropriate model architecture for the given task and dataset.

In comparison to the first method in this area, ST-Net~\cite{He2020}, our BrST-Net framework improved the gene expression prediction performance quite substantially. We have completely trained the most recent state-of-the-art CNN models and transformers and compared their performances on the ST dataset, while ST-Net employed a DenseNet trained on natural images (such as cats, dogs, and flowers), which may not be optimal for histopathology. In our suggested approach, in addition to the core network, an auxiliary network is added to predict the remaining genes. The auxiliary loss helps reduce the fading gradients problem and stabilizes and regularises the training. Whereas ST-Net can predict only 102 genes with positive correlation, our framework was able to predict 237. ST-Net revealed the top-5 predicted genes with a median correlation coefficient value of (0.34, 0.33, 0.31, 0.30, and 0.29), and for smoothed data (0.49, 0.50, 0.50, 0.52, and 0.43). In contrast, with our framework using EfficientNet-b0 with the proposed AuxNet, we could predict 24 genes with a median correlation coefficient greater than 0.50, which represents a considerable increase in performance.

Notwithstanding the merits of BrST-Net, there is still much room for further study and development in this field. As is well known, deep learning methods require a large dataset to obtain good results. Thus we expect future studies using larger datasets to increase the prediction performance and robustness of the models. A richer dataset from a larger number of patients would also expand our framework's generalisability. Because each model has unique performance characteristics, combining two or more models can be beneficial. Also, it may be possible to do further gene expression filtering and model training for predicting target gene expression. There is a need to further improve the number of genes predicted with high correlation, which are of established relevance in cancer biology and treatment. In the meantime, BrST-Net could serve as an inexpensive and fast high-throughput screening tool for large numbers of patient samples to direct downstream definitive molecular analyses.

\section*{Acknowledgements}
EKAM is supported by a Researcher Exchange and Development in Industry (REDI) Fellowship from MTPConnect and MRFF Australia. The authors thank the National Computational Infrastructure (NCI) Australia for providing access to the Gadi high-performance computing system, which contributed to the research results reported in this paper.

\bibliographystyle{unsrt}
\bibliography{paper}

\newpage
\section*{Supplementary Materials}
\beginsupplement

\begin{table*}[!h]
\centering
\caption{Top 10 predicted genes by our framework using different models without AuxNet. The numbers are median Pcc values on the heldout test patient ID BC23903. Here, to save space, ``EfficientNet'' is abbreviated to ``ENet'', and ``Inception-v3'' to ``Incep-v3''. Bold indicates best performance (highest Pcc) per gene and underlined indicates second-best.}
\vspace{1em}
\resizebox{1\textwidth}{!}{
\begin{tabular}{lccccccccccc}
\hline
\textbf{Genes} & \textbf{ResNet101} & \textbf{Incep-v3} & \textbf{ENet-b0} & \textbf{ENet-b1} & \textbf{ENet-b2} & \textbf{ENet-b3} & \textbf{ENet-b4} & \textbf{ENet-b5} & \textbf{ViT-B16} & \textbf{ViT-B32} \\
\hline
B2M & 0.5715 & 0.5685 & 0.5746 & 0.5695 & 0.5616 & 0.5497 & 0.5798 & \textbf{0.5922} & \underline{0.5887} & 0.4698 \\
ACTG1 & 0.5088 & 0.5480 & 0.6087 & 0.5703 & \underline{0.6090} & 0.5305 & 0.5909 & \textbf{0.6142} & 0.5468 & 0.4414 \\
ACTB & 0.5396 & 0.5385 & 0.5171 & 0.5293 & 0.5811 & \textbf{0.5865} & 0.4980 & \underline{0.5859} & 0.5290 & 0.4470 \\
TMSB10 & 0.4410 & 0.4018 & 0.3313 & 0.3662 & 0.5438 & \underline{0.5679} & 0.4717 & \textbf{0.6063} & 0.1378 & 0.3972 \\
GNAS & 0.5237 & 0.5613 & \textbf{0.5952} & 0.5581 & 0.5541 & 0.5089 & 0.5654 & \underline{0.5819} & 0.5093 & 0.5038 \\
PTMA & 0.5402 & 0.5099 & 0.5155 & 0.5332 & \textbf{0.5530} & 0.5229 & 0.5305 & \underline{0.5494} & 0.4796 & 0.4047 \\
PTPRF & 0.4503 & 0.4852 & \underline{0.5327} & 0.4970 & 0.5171 & 0.5111 & 0.4957 & \textbf{0.5531} & 0.4404 & 0.3742 \\
ERBB2 & 0.4635 & 0.3730 & 0.4977 & 0.4494 & 0.5406 & \underline{0.5414} & 0.5391 & \textbf{0.5502} & 0.4056 & 0.3994 \\
PRDX1 & 0.4364 & 0.4544 & 0.4392 & 0.3734 & \underline{0.5009} & 0.4826 & 0.4954 & \textbf{0.5091} & 0.4538 & 0.4245 \\
TMSB4X & 0.4544 & 0.4139 & 0.4491 & 0.4259 & 0.5491 & 0.5072 & \underline{0.5523} & \textbf{0.5674} & 0.3665 & 0.4450 \\
\hline
\end{tabular}
}
\label{SupplementaryTableS1}
\end{table*}

\begin{table*}[!h]
\centering
\caption{Results of the cross-validation (mean +/- stdev over the 5 folds) with AuxNet.}
\vspace{0.5em}
\begin{tabular}{l@{\hspace{3em}}c@{\hspace{2em}}c}
\hline
\multicolumn{1}{l}{\multirow{2}{*}{\textbf{Model}}} & \multicolumn{2}{c}{\textbf{Validation data}} \\ \cline{2-3} 
\multicolumn{1}{c}{}                       & \multicolumn{1}{c}{\textbf{aMAE}} & \multicolumn{1}{c}{\textbf{aRMSE}} \\ \hline
ResNet101 + AuxNet                         & 0.9001 $\pm$ 0.0417         & 1.1473 $\pm$ 0.0621          \\
Inception-v3 + AuxNet                      & 0.8629 $\pm$ 0.0713         & 1.0612 $\pm$ 0.0905          \\
EfficientNet-b0 + AuxNet                   & 0.8274 $\pm$ 0.0469         & 1.0405 $\pm$ 0.0762          \\
EfficientNet-b1 + AuxNet                   & 0.8287 $\pm$ 0.0409         & 1.0375 $\pm$ 0.0554          \\
EfficientNet-b2 + AuxNet                   & 0.8452 $\pm$ 0.0667         & 1.0516 $\pm$ 0.0493          \\
EfficientNet-b3 + AuxNet                   & 0.8674 $\pm$ 0.0617         & 1.1021 $\pm$ 0.1575          \\
EfficientNet-b4 + AuxNet                   & 0.9334 $\pm$ 0.1554         & 1.3574 $\pm$ 0.4783          \\
EfficientNet-b5 + AuxNet                   & 0.8588 $\pm$ 0.0823         & 1.1314 $\pm$ 0.1856          \\
ViT-B16 + AuxNet                           & 0.8588 $\pm$ 0.0520         & 1.0534 $\pm$ 0.0686          \\
ViT-B32 + AuxNet                           & 0.8783 $\pm$ 0.0639         & 1.1386 $\pm$ 0.0759          \\ \hline
\end{tabular}
\label{SupplementaryTableS2}
\end{table*}

\begin{table*}[!h]
\centering
\caption{Results of the cross-validation (mean +/- stdev over the 5 folds) without AuxNet.}
\vspace{0.5em}
\begin{tabular}{l@{\hspace{3em}}c@{\hspace{2em}}c}
\hline
\multicolumn{1}{l}{\multirow{2}{*}{\textbf{Model}}} & \multicolumn{2}{c}{\textbf{Validation data}} \\ \cline{2-3} 
\multicolumn{1}{c}{}                       & \multicolumn{1}{c}{\textbf{aMAE}} & \multicolumn{1}{c}{\textbf{aRMSE}} \\ \hline
ResNet101                                  & 0.8489 $\pm$ 0.0624         & 1.0388 $\pm$ 0.0696          \\
Inception-v3                               & 0.8749 $\pm$ 0.0538         & 1.0815 $\pm$ 0.0634          \\
EfficientNet-b0                            & 0.8326 $\pm$ 0.0867         & 1.0260 $\pm$ 0.1009          \\
EfficientNet-b1                            & 0.8854 $\pm$ 0.0562         & 1.0880 $\pm$ 0.0683           \\
EfficientNet-b2                            & 0.8485 $\pm$ 0.0854         & 1.0446 $\pm$ 0.1022          \\
EfficientNet-b3                            & 0.8868 $\pm$ 0.0568         & 1.0876 $\pm$ 0.0695          \\
EfficientNet-b4                            & 0.8309 $\pm$ 0.0574         & 1.0514 $\pm$ 0.0989           \\
EfficientNet-b5                            & 0.9165 $\pm$ 0.0676         & 1.3785 $\pm$ 0.5747          \\
ViT-B16                                    & 0.8370 $\pm$ 0.1193         & 1.0468 $\pm$ 0.1037          \\
ViT-B32                                    & 0.8860 $\pm$ 0.0524         & 1.0852 $\pm$ 0.0712          \\ \hline
\end{tabular}
\label{SupplementaryTableS3}
\end{table*}

\begin{figure*}[!t]
\centering
\includegraphics[width=\textwidth]{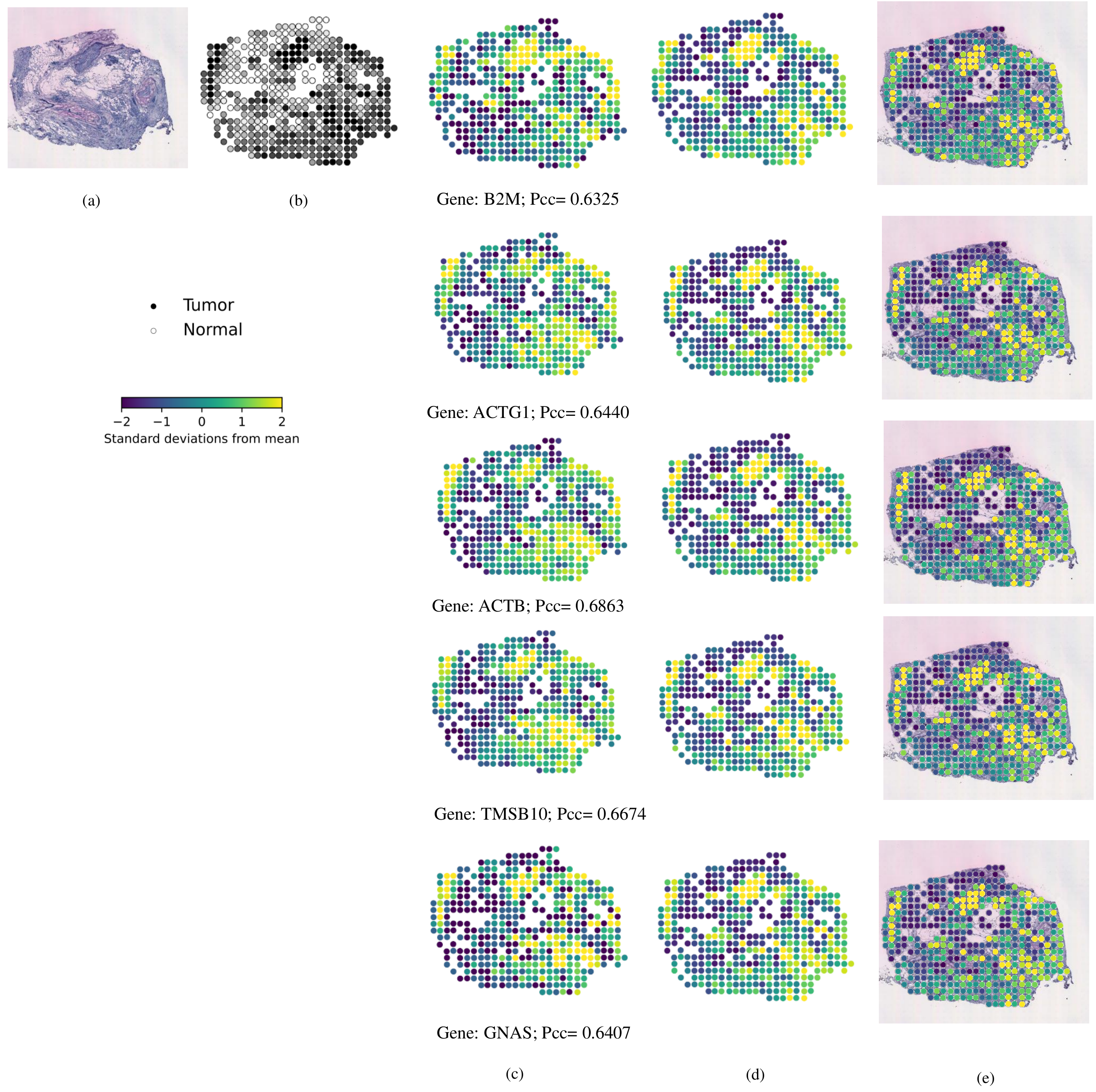}
\caption{Visualisation of gene expression for patient BC23903 slide number C2. (a) Histopathology sample from the test data. (b) Binary label of tumour (black) and normal (white) areas, with gray parts indicating areas that are not purely tumor or normal. (c) Ground truth expression of corresponding genes. (d) Predicted expression of corresponding genes and their Pcc values (e) Visualisation of predicted gene expression overlayed on the tissue slice.}
\label{sv_1}
\end{figure*}

\begin{figure*}[!t]
\centering
\includegraphics[width=\textwidth]{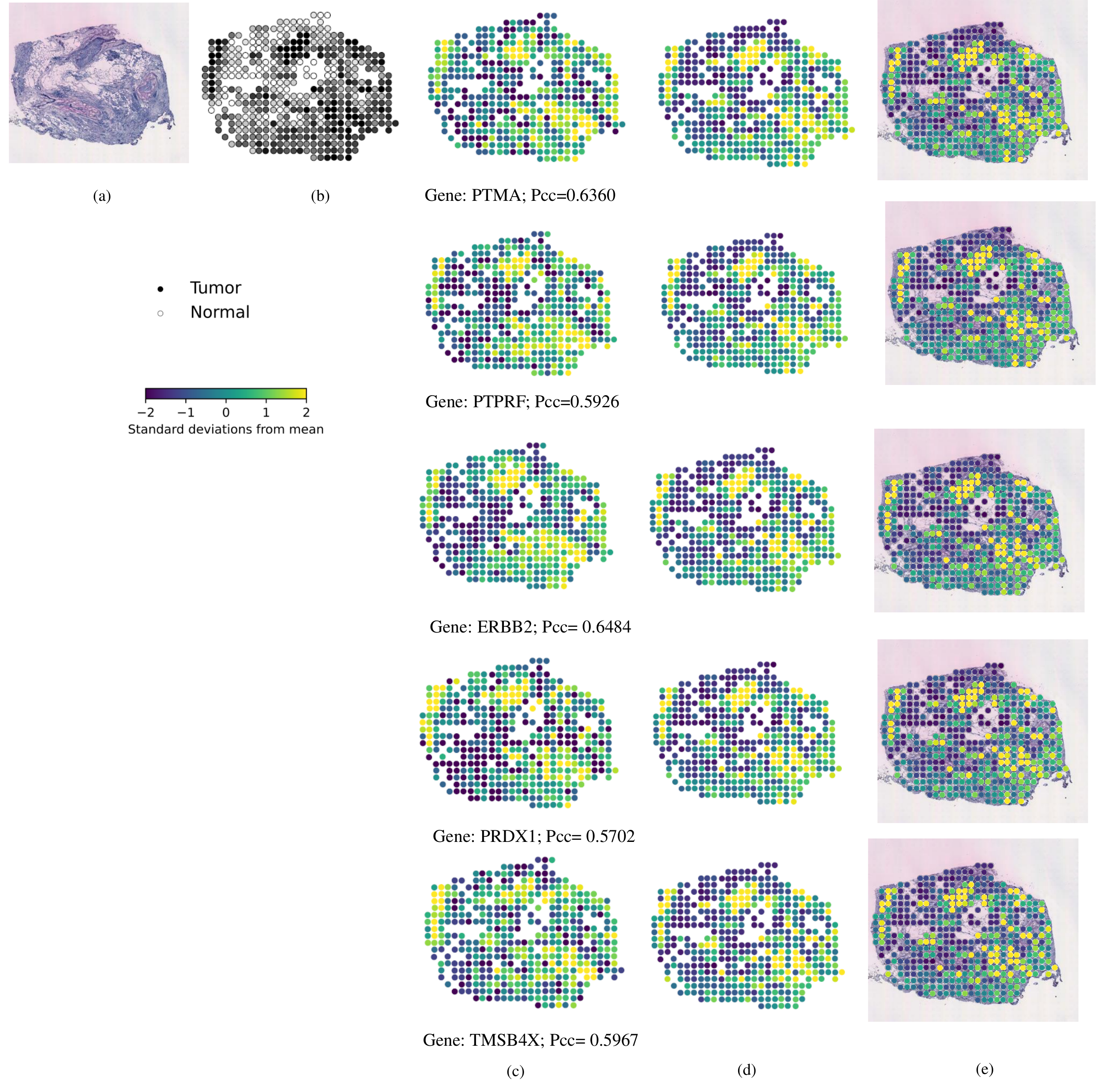}
\caption{Visualisation of gene expression for patient BC23903 slide number C2. (a) Histopathology sample from the test data. (b) Binary label of tumour (black) and normal (white) areas, with gray parts indicating areas that are not purely tumor or normal. (c) Ground truth expression of corresponding genes. (d) Predicted expression of corresponding genes and their Pcc values (e) Visualisation of predicted gene expression overlayed on the tissue slice.}
\label{sv_2}
\end{figure*}

\end{document}